\documentclass[12pt]{article}
\begin{document}
\title{Retroduction, Multiverse Hypotheses and Their Testability*}
\author{William R. Stoeger, S. J.**}
\maketitle
\noindent
*Based on a talk given at the symposium ``Multiverse and String
Theory: Toward Ultimate Explanations in Cosmology,'' held on
19-21 March 2005 at Stanford University.\\

\noindent
**Vatican Observatory Research Group, Steward Observatory, The
University of Arizona, Tucson, Arizona 85721

\begin{abstract}
The actual existence of collections of universes -- multiverses --
is strongly suggested by leading approaches to quantum cosmology,
and has been proposed earlier as an attractive way to explain 
the apparent fine-tuned character of our universe. But, how
can such hypotheses be tested? After briefly discussing the
key distinction between possible and really existing
multiverses, and the importance of an adequate generating
process, we focus on elaborating how multiverse hypotheses
can be retroductively tested, even though they will probably
never be directly observed. In this approach, scientific acceptance
of multiverses would rely on the long-term success and
fertility of quantum cosmological theories including them as
essential elements or as inevitable consequences.  
\end{abstract}

\section{Introduction}
As we struggle to understand our universe more fully, using all the
resources of physics, astronomy, cosmology and even philosophy, we
are finding more and more indications that it may very well be
just one of a very large number of universes, or universe domains.
We often refer to such a collection, or ensemble, of universes
as a ``multiverse.'' If these preliminary indications are correct, the
multiverse to which our own universe belongs would have emerged from
some quantum cosmological process that, from our present very limited
perspective, is hidden in what we often refer to as the Big Bang. In other words,
the more we delve into trying to understand how our own universe, or our
own region of a much larger universe, was generated, the more we encounter
the likelihood that whatever primordial process was involved generated a
very large number of other universes, or universe domains. \\

 The emergence
of such multiverses was first proposed and discussed in detail by Linde (1982,
1983a). In doing so he emphasized that a multiverse as such is a collection of
different universes, or one universe consisting of different large regions, 
representing a wide-range of different properties and different low-energy laws
of physics (in particular, very different values of the gravitational,
electromagnetic and nuclear coupling constants, and very different masses of the fundamental 
particles). Thus, as he then proposed, it provides a way of explaining the apparent fine-tuning of our
universe for complexity (``the anthropic principle''). About the same time, Vilenkin (1983)
described in a similar vein the creation of many universes by inflationary processes, without
referring to their possible application to resolving the fine-tuning issue. Soon afterwards 
Linde (1983b, 1990) developed his multiverse idea much more fully in his chaotic inflationary scenario. 
Basically, Linde envisions a primordial cluster of tiny, causally
separate regions as physical reality emerges from the Planck era, where quantum
gravity dominates. Each of these regions becomes a separate universe or universe
domain -- some inflating, and some not, depending on the value of the fluctuating
scalar fields in a given region as the transition to classical space-time is
negotiated. Besides Vilenkin's and Linde's work, many others (Sciama 1993, Leslie 1996, Deutsch 1998,
Tegmark 1998, 2003, Smolin 1999, Weinberg 2000, Lewis 2000, Rees 2001) have since
discussed general ways in which an ensemble of universes might originate.
More recently superstring theory has given more specific impetus to the multiverse
idea. Versions of it provide ``landscapes'' populated by extremely large numbers
of vacua, each of which could initiate a separate universe domain (Kachru, et al.
2003; Susskind 2003; Freivogel and Susskind 2004; Freivogel, et al. 2005; the other articles in this
volume, and references therein). \\

The recognition that our universe appears to be fine-tuned
for complexity, and for life and consciousness (Dicke 1961, Collins and Hawking 1973,
Carter 1974, McMullin 1993) has actually been an independent and earlier stimulus to
considering multiverses. If any one of a number of
key constants, or other parameters, describing our universe and its dynamics
had slightly different values, our universe would be so different that it 
would not support chemically complex systems. Therefore, it would be forever without
the possibility of life.  This has become
known as the ``anthropic principle.'' What accounts for this apparent delicate
adjustment of constants and parameters? Why do they have the values they have,
rather than other values? The existence of a large collection of universes, 
which represents a significant range of possible cosmic parameter values and to which
our own universe belongs, would be a possible scientifically acceptable way of
explaining such apparent fine-tuning -- even though it would not provide
an ultimate philosophical explanation. This solution was first suggested by
Collins and Hawking (1973), and has since become the predominant and really
only scientific proposal for resolving the anthropic enigma (see McMullin 1993, Carr
2006, and the other articles in this volume, and references therein). If quantum
cosmological processes naturally produced a large variety of universes, including
ours, then we would simply find ourselves in one in which all the many conditions
for life and consciousness have been fulfilled. This would be similar to how we
might explain the very special conditions we enjoy on Earth and in our
Solar System -- our Sun being one of several hundred billion stars
in the Milky Way. \\

Of course, the multiverse answer to the fine-tuning puzzle is scientifically
acceptable only if multiverses themselves are really legitimate objects of scientific
inquiry! Are they? Some have argued that, since we shall never be able to
directly detect multiverses or make observations of them, they fall outside
the realm of scientific investigation (Gardner 2003). This important question
leads us directly to the focus of this chapter, the testability of multiverse
hypotheses as a philosophy of science issue. In other words testability is
a necessary condition for scientific legitimacy. Relying on the fundamental 
insights of the American philosopher C. S. Peirce concerning ``retroduction,''
and its development and historical confirmation as ``the inference which makes science,''
by Ernan McMullin (1992), I argue that multiverse hypotheses are
scientifically testable. Then, I shall briefly outline what general {\it scientific}
requirements must be met to satisfy the philosophical standards of retroductive
testabililty. Before tackling the specific issue of testability itself, I shall
briefly address the connected issues of the difference between possible and
existing ensembles of universes -- possible and really existing multiverses --
and the need for a definite generating process for any existing multiverse.
\footnote {There
are other important and fascinating philosophical issues raised by the
possibility of multiverses, such as those of realized infinity (can we really
have an infinity of really existing universes?), ontological and causal
reductionism (can new qualities emerge in the course of the evolution of
a given universe, or all possible emergent qualities simply latent in the 
physics from the beginning?), and that of the choice between the generic
and the special (can we completely avoid fine-tuning and/or special
``initial'' conditions?). But the issue of testability, and the closely
connected ones of possible and existing multiverses, and the need for
a physical generating mechanism, are the most urgent for multiverse cosmology at present.} Many
of the points I shall emphasize in this regard may seem obvious or trivial, but they are 
crucial in providing a secure basis for our
consideration of testability, as well as of other related physical and
philosophical issues.

\section{Possible and Existing Multiverses, and Generating Processes}

From both a physical and a philosophical point of view, it
is crucial to distinguish between the ensemble of all
possible universes or universe domains, and any ensemble of
existing universes (Ellis, {\it et al.} 2003, Stoeger, {\it et al.}
2004). Though we may conceive all possible
universes existing, philosophically, it is almost certainly the case that only a subset of
these actually exists. Secondly, it is obvious that
an existing ensemble of universes or universe domains requires some process or
series of processes to produce them -- to actualize them. This
is one of the challenges of quantum cosmology -- to determine
the physics of the primordial process by which our universe was born, and the
collection of universes or universe domains, and their specific properties,
which have emerged in association with it. It is the set of all
existing universes which needs to be explained by cosmology and
physics, {\it not} the ensemble of possible universes. Finally,
it is only the set of really existing universes which would provide
an adequate answer to the fine-tuning problem (McMullin 1993, p. 371; Ellis, {\it et al.}
2003; Stoeger, {\it et al.} 2004). \\

Though we cannot adequately describe the space ${\cal M}$ of possible universes $m$,
we can set up a heuristic classical framework for doing so (see Ellis, {\it et al.},
2003), based on different reasonable assumed sets of laws of nature -- either laws of physics or meta-laws
that determine the laws of physics -- the general parameter classes of which all $m$ have in common.
\footnote {Such a classical, non-quantum-cosmological description of ${\cal M}$ is provisional,
not fundamental. It provides us with a preliminary systematic framework, consistent with
our present limited understanding of cosmology, within which to begin studying multiverses. As
quantum cosmology matures, we shall have to develop a more fundamental quantum framework which
takes such issues as quantum entanglement and decoherence into consideration. It may very well be that
as ``the wave function of the universe'' decoheres, an entire ensemble of universes emerges. 
These would all be entangled with one another, and ``the wave function of the universe'' would
provide the fundamental basis for the quantum ontology of the multiverse, as well as the
seed from which it was generated. Then, we might very well want to consider the meta-structure
of the set of all possible ``waves functions of the universe.''  However, at present we do not
have an adequate theory of quantum gravity, or of quantum cosmology, to enable us to proceed
meaningfully in accomplishing these two tasks.} Without this we have
no basis for defining ${\cal M}$. In general we must incorporate in ${\cal M}$ at least the geometry
of the allowed universes and the physics of the matter they contain. Clearly, we do not
have any way of reliably determining the contours of what is really possible. However, on the basis
of what does not contradict philosophical or sensible physical principles, we can 
set up a parameter space for what we presently consider possible universes (Ellis, {\it et al.}
2003; Stoeger, {\it et al.} 2004). \\

In any really existing ensemble of universes or universe domains many of the
universes in $\cal{M}$ will not be realized, and some may be realized more
than once. Essentially, the physics of the primordial vacuum, or other
fundamental configuration, out of which a given
multiverse emerges, together with that of the generating processes operating
to produce it, will effectively determine a distribution function $f(m)$ specifying
how many times each $m$ in $\cal M$ is realized (Ellis, {\it et al.} 2003). This expresses the contingency
of any multiverse actualization -- the fact that not every possible universe
has to be realized -- and the detailed physical dependence of the existing
multiverse on the underlying physics, whatever that may be. \footnote {There are other
technical issues related to this, of course, including whether or not we can define a
unique measure on the $\cal M$ and on its subset of realized universes.} \\
 
Two almost trivial implications follow from this. The first is that
there is no unique really existing multiverse -- there are an infinite number of
possible such multiverses (that is distribution functions $f(m)$ which could be
defined on $\cal M$). Which is the one to which our observable universe
belongs depends, as I have stressed, on the primordial physics that is
operative. The second implication is that, though a knowledge of the
primordial physics would give us an account of the beginning of our
universe and of the multiverse to which it belongs, as well as of
the apparently fine-tuned character of our universe, it would raise the
further even deeper question of its own origins, and whether or not it
and the processes it governs require fine-tuning. Here an infinite regress
lurks in the wings. From a philosophical point of view, then, any such   
primordial quantum physics, though making our universe more intelligible
from a scientific point of view, would not provide an ultimate explanation for it
nor account for its ultimate origin. We would still be able ask with good 
reason for an explanation of the primordial vacuum and for the physics
which governs it.  \\

We have been discussing the need for a definite generating process for
any existing multiverse. That requires some definite, detailed quantum-gravity
physics, which in turn would have to be shown to determine the overall distribution $f(m)$,
and the range of properties represented by $f(m)$. Obviously, at least one of those
universes would have to have the properties of our universe. This is very a demanding
requirement! As we now know from considering the apparent fine-tuned character of
our universe, there are very narrow ranges of values of the fundamental constants, and
of other cosmic parameters (e. g., the density of dark energy), that allow for chemical
complexity and for life. Outside of these ranges, complexity, and life, would not be
possible. A little reflection also assures us, besides, that it is the underlying physics of the 
multiverse generating process which induces certain common features of structure and content
in all the universes it produces, despite their great variety. Thus, it will be the basic
link connecting all the universes in the ensemble. Through that generating process they
will all be governed by a common set of fundamental primordial laws, or meta-laws. Otherwise,
there would be no reason to consider the universes to be members of the same multiverse, nor
any way of relating them to one another. Their common origin in a specific physical process, or 
chain of processes, governed by a common physics in the only way of achieving that. \\ 

Finally, in this regard, it is helpful to realize that there may be, as has been
often mentioned, innumerable multiverses which
constitute reality. In light of what we have just been discussing, all but our own multiverse would, by definition, be
causally and generationally disconnected from us. Thus, we would
evidently have no possible way of discovering their
existence, nor any reason to postulate their contribution to the
intelligibility of our universe (Stoeger, {\it et al.} 2004; Stoeger 2004).
The only way that would be
conceivable is if the generating processes of those multiverses
were associated with those of ours. But in that case they would really be
part of our multiverse. \\ 
  
\section{Retroduction} 

From what I have just pointed out, it is fairly clear that
really existing multiverses completely disassociated from
ours will not be subject to any scientific confirmation. We
will never be able to rule them out, but we shall never be
able to present positive direct or indirect evidence for
their existence. But how about for the existence and
character of our own multiverse -- for universes or universe
domains primordially connected in some way to ours? It
seems that, if they are connected in some way to our universe,
e. g. by a common generating process or an initial vacuum
state, then there should be in principle ways to find out
about them. \\

There may, in fact, even be relatively direct ways, as 
pointed out by Freivogel and Susskind (2004), who 
demonstrate that the bubble universes out beyond our
horizon are not, according string theory models, completely decoupled. Our horizon 
will contain information, scrambled though it be, concerning
the other universes which exist beyond it and to which
it is generationally linked. However, according to some
experts in that field (Susskind, private communication), recovering
such information would probably not be feasible -- it would take
an enormously long time. \\

There is an attractive and compelling approach to 
scientific testability which would enable
us to indirectly establish the existence of our
multiverse, and its characteristics, in much more
promising way -- under certain
well specified conditions. This brings us to a
consideration of C. S. Peirce's concept of 
of ``retroduction,'' or ``abduction,'' which has
been rather compellingly argued by Ernan McMullin
(1992) as ``the inference which makes science.'' Retroduction,
according to McMullin, is 
the rational process by which scientific conclusions
are most often and most fruitfully reached. \\

How does retroduction function? On the basis of what
researchers know, they construct imaginative hypotheses,
which are then used to probe and to describe the 
phenomena in deeper and more adequate ways than before.
As they do so, they will modify or even replace the
original hypotheses, in order to make them more fruitful
and more precise in what they reveal and explain. The
hypotheses themselves may often presume or directly
imply the existence of
certain hidden properties or entities (like multiverses!)
which are fundamental to or consequent upon the explanatory power they possess.
As these hypotheses become more and more fruitful in
revealing and explaining the natural phenomena they
investigate and their inter-relationships (rendering them
more and more intelligible), and more central to scientific
research in a given discipline, they become more and more
reliable accounts of the reality they purport to model or
describe. Even if some of the hidden properties or entities
they postulate are never directly detected or observed, the 
long term success and fruitfulness of the hypotheses 
indirectly leads us to affirm that something like them probably 
exists. \\

Thus, from this point of view, the existence of an ensemble of
universes or universe domains would be strongly, though provisionally, supported
(but not logically deduced!).  The fundamental retroductive requirement is that
the existence of the multiverse is a key component or consequence of
hypotheses which are successful and fruitful
in the long term. In other words, as a basic, though observationally
inaccessible component of the theory, it provides greater intelligibility and
understanding of our universe. The hypotheses themselves enable us to make testable predictions 
which, if fulfilled, provide a more thorough and more coherent explanation
of cosmic phenomena we observe than competing theories do.  If such
indirect support is not forthcoming, then all we can do is to treat 
the multiverse hypothesis as a promising speculative scenario needing
further development and requiring further fruitful application. That is 
where we are now, I believe. But there are definite prospects for
improving our confidence in it. \\

At this point we should briefly indicate how we are to judge 
``long-term success and fruitfulness'' of a given set of 
hypotheses. In general, a theory is fruitful and qualitatively
and quantitatively supported if it (McMullin 1992; also, see Allen
2001): 1.) accounts for all the
relevant data (empirical adequacy); 2.) provides a {\it continuing}
foundation for explanatory success, and stimulates further fruitful
investigation (theory fertility); 3.) establishes the compatibility
of previously disparate domains of observed phenomena (unifying power);
4. manifests consistency (or correlation) with other established 
theories (theoretical coherence). These are the broad criteria
we need to apply to specific theories predicting the existence of a multiverse.
We now briefly discuss what that involves.

\section{Retroductive Testability of the Multiverse Hypothesis}

Relying on these insights concerning retroduction, then, we could make 
a reasonable claim for the existence of a multiverse, if we could
show that its existence was a more or less inevitable
consequence of well-established physical laws and processes. This is essentially
the claim that is made for chaotic inflation (Linde 1990). However, the 
challenge is that the proposed underlying physics has not been adequately
tested, and may be untestable. We need evidence that the postulated physics --
particularly that governing the quantum cosmological generating processes --
is true in this universe. \\

Thus, there are two further basic requirements which must still be met, once we
have proposed a viable ensemble or multiverse theory. The first is to
provide some credible link between these vast extrapolations from
presently known physics to the physics in which we have some confidence.
The second is to provide some at least indirect evidence that the scalar
potentials, or other overarching cosmic principles central to generating
bubble universes (e. g., a superstring
theory of a given type), really have been
functioning in the very early universe, or before its emergence. \\

The issue is not just that the inflaton has not been identified and its
potential is untested. It is also that, for example, we are assuming
quantum field theory remains valid far beyond the domain where it has
been tested, especially given the unsolved problems at the foundations
of quantum theory, the divergences of quantum field theory, and 
failure of the theory to resolve the cosmological constant problem. \\

Once these basic requirements are met, then the theory involving 
multiverses must, as it is further developed, continue to receive
indirect support from further theoretical and observational work,
lead to new promising lines of inquiry which open up and enrich
cosmology, and contribute substantially to the overall coherence
and unity of physics and cosmology. In short, it must, over time,
continue to provide a reliable foundation for increasing our
understanding of our universe, its origin and its characteristics. \\

In conclusion, despite the rigor that it demands,  the retroductive
approach to scientific testability does open the way for scientific
confirmation of the existence of a multiverse to which our universe
belongs, and thus to a scientific resolution to the fine-tuning
problem. Thus, at least potentially, the issues connected with
multiverses {\it can} be brought in from the realms of pure
metaphysics to those where scientific confirmation is possible. \\  

\noindent
{\Large \bf Acknowledgements} \\
Special thanks to the John Templeton Foundation who sponsored the symposium,
``Multiverse and String Theory: Toward Ultimate Explanations in Cosmology,''
at which this paper was originally presented in March 2005, and who have
encouraged us to publish our work in this volume; to George Ellis,
my research partner in investigating the physics and philosophy of multiverses;
and to the participants of the symposium, especially Andrei Linde, who encouraged me in thinking along
these lines, and from whom I have learned so much. \\
  
\noindent
{\Large \bf References} \\

\noindent
Allen, Paul, 2001, {\it Philosophical Framework with the Science-Theology Dialogue:
A Critical Reflection on the Work of Ernan McMullin}, unpublished Ph.D dissertation,
St. Paul University, Ottawa, Canada. \\

\noindent
Carr, Bernard, editor, 2006, {\it Universe or Multiverse},
Cambridge University Press, in press. \\

\noindent
Carter, B., 1974, ``Large Number Coincidences and the Anthropic
Principle in Cosmology,'' in M. S. Longaire, editor, {\it
Confrontation of Cosmological Theory with Astronomical Data},
Dordrecht, Reidel, pp. 291-298. \\

\noindent
Collins, C. B. and Hawking, S. W., 1973, ``Why is the Universe
Isotropic?'' {\it Astrophys. J.}, 180, 317-334. \\

\noindent
Deutsch, D. 1998, {\it The Fabric of Reality: The Science of 
Parallel Universes -- and its Implications} (Penguin).\\

\noindent
Dicke, R. H., 1961, ``Dirac's Cosmology and Mach's Principle,''
{\it Nature} 192, 440-441. \\

\noindent
Ellis, G. F. R., Kirchner U. and Stoeger, W. R., 2003, ``Multiverses
and Physical Cosmology,'' {\it Mon. Not. R. astr. Soc.}, 347, 921. \\

\noindent
Freivogel, B. and Susskind, L. 2004, ``Framework for the
String Theory Landscape,'' {\it Phys. Rev.}, D70, 126007. \\

\noindent
Freivogel, B., Kelban, M., Rodr\'{i}guez Mart\'{i}nez, and Susskind,
L., 2005, ``Observational Consequences of a Landscape,''
arXiv: hep-th/0505232v2. \\

\noindent
Gardner, M., 2003, {\it Are Universes Thicker than Blackberries?},
New York, Norton. \\

\noindent
Kachru, S., Kallosh, R., Linde, A., and Trivedi, S. P., 2003,
{\it Phys. Rev.}, D68, 046005. \\
 
\noindent
Leslie, J. 1996, {\it Universes} (Routledge). \\

\noindent
Lewis, D. K. 2000, {\it The Plurality of Worlds} (Oxford: 
Blackwell). \\

\noindent
Linde, A. D. 1982, ``Nonsingular Regenerating Inflationary 
Universe,'' Cambridge University Preprint-82-0554;
http://www.stanford.edu/~alinde/1982.pdf. \\

\noindent
Linde, A. D. 1983a, ``The New Inflationary Universe Scenario,''
in {\it The Very Early Universe}, ed. by G. W. Gibbons,
S. W. Hawking and S. Siklos, Cambridge University Press, 
pp. 205-249. \\

\noindent
Linde, A. D. 1983b, {\it Phys. Lett.}, B129, 177. \\

\noindent
Linde, A. D. 1990, {\it Particle Physics and Inflationary
Cosmology} (Chur: Harwood Academic Publishers). \\

\noindent
McMullin, E. 1993, ``Indifference Principle and Anthropic
Principle in Cosmology,'' {\it Stud. Hist. Phil. Sci.},
24, 359-389. \\

\noindent
McMullin, E. 1992, {\it The Inference that Makes Science}
(Milwaukee, WI: Marquette University Press). \\

\noindent
Rees, M. J. 2001, {\it Our Cosmic Habitat} (Princeton:
Princeton University Press). \\

\noindent
Sciama, D. 1993, {\it Is the Universe Unique?}, in
{\it Die Kosmologie der Gegenwart}, ed. G. B\"{o}rner
and J. Ehlers (Serie Piper). \\

\noindent
Smolin, L. 1999, {\it The Life of the Universe} (Oxford:
Oxford University Press). \\

\noindent
Stoeger, W. R. 2004, ``What is `the Universe' which Cosmology
Studies?'' in {\it Fifty Years in Science and Religion: Ian G.
Barbour and His Legacy}, Robert John Russell, editor (Aldershot,
U. K. and Burlington, VT: Ashgate Publishing Co.), pp. 127-143. \\

\noindent
Stoeger, W. R., Ellis, G. F. R., and Kirchner, U. 2004,
``Multiverses and Cosmology: Philosophical Issues,''
arXiv: astro-ph/0407329v2. \\

\noindent
Susskind, L. 2003, ``The Anthropic Landscape of String Theory,''
arXiv:hep-th/0302219v1. \\

\noindent
Tegmark, M., 1998, ``Is the Theory of Everything Merely the Ultimate
Ensemble Theory?'' {\it Annal. Phys.}, 270, 1-51. \\

\noindent
Tegmark, M. 2003, ``Parallel Universes,'' {\it Scientific American}
(May 2003), 41-51. \\

\noindent
Vilenkin, A. 1983, ``The Birth of Inflationary Universes,''
{\it Phys. Rev.}, D27, 2848. \\

\noindent
Weinberg, S. 2000, ``A Priori Probability Distribution of the
Cosmological Constant,'' {\it Phys. Rev.}, D61, 103505.\\ 
\end{document}